\documentstyle[aps,prb,preprint,floats]{revtex}
\input epsf

\begin{document}
\draft
\title{Infrared conductivity of a \bbox{d_{x^2-y^2}}-wave
superconductor with impurity and spin-fluctuation scattering}
\author{S. M. Quinlan}
\address{Solid State Division, Oak Ridge National Laboratory,
P.O. Box 2008, Oak Ridge, TN 37831--6032 and
Department of Physics and Astronomy, University of Tennessee,
Knoxville, TN 37996--1200}
\author{P. J. Hirschfeld}
\address{Department of Physics, University of Florida,
Gainesville, FL 32611}
\author{D. J. Scalapino}
\address{Department of Physics,
University of California,
Santa Barbara, CA 93106--9530}
\date{October 16, 1995}
\maketitle
\begin{abstract}
Calculations are presented of the in-plane far-infrared conductivity
of a $d_{x^2-y^2}$-wave superconductor, incorporating elastic scattering due
to impurities and inelastic scattering due to spin fluctuations.  The
impurity scattering is modeled by short-range potential scattering
with arbitrary phase shift, while scattering due to spin fluctuations
is calculated within a weak-coupling Hubbard model picture.  The
conductivity is characterized by a low-temperature residual Drude
feature whose height and weight are controlled by impurity scattering,
as well as a broad peak centered at $4\,\Delta_0$ arising from
clean-limit inelastic processes.  Results are in qualitative agreement
with experiment despite missing spectral weight at high
energies.
\end{abstract}

\narrowtext

\section{Introduction}

Infrared studies of the classic low-temperature superconductors
provided evidence for the existence of a superconducting energy gap as
well as information on the phonon-mediated pairing
interaction.\cite{Nam,Brandi} In particular, at a low reduced
temperature $T/T_c$, the conductivity $\sigma_1(\omega)$ of an $s$-wave
superconductor shows an onset when $\omega$ exceeds $2\,\Delta$,
increasing to the normal-state value at frequencies several times
$2\,\Delta$.\cite{Bickers} Further, structure in $\sigma_1(\omega)$ at
$2\,\Delta+\omega_p$ reflects peaks at $\omega_p$ in the effective
phonon mediated interaction $\alpha^2F(\omega)$ and a detailed
measurement of $\sigma_1(\omega)$ can in principle\cite{Brandi} be used to
determine $\alpha^2F(\omega)$.  Thus it was hoped that infrared
measurements of $\sigma_1(\omega)$ for the high-temperature superconducting
cuprates would provide similar information on both the gap and the
pairing mechanism.  However, the search for evidence of an energy gap
in the cuprates has proven difficult.\cite{Tanner} In spite of a
confluence of data on high-quality samples, there has been a wide,
even divergent, view on the interpretation of these data.

\begin{figure}
\epsfysize=6.5in
\epsfbox{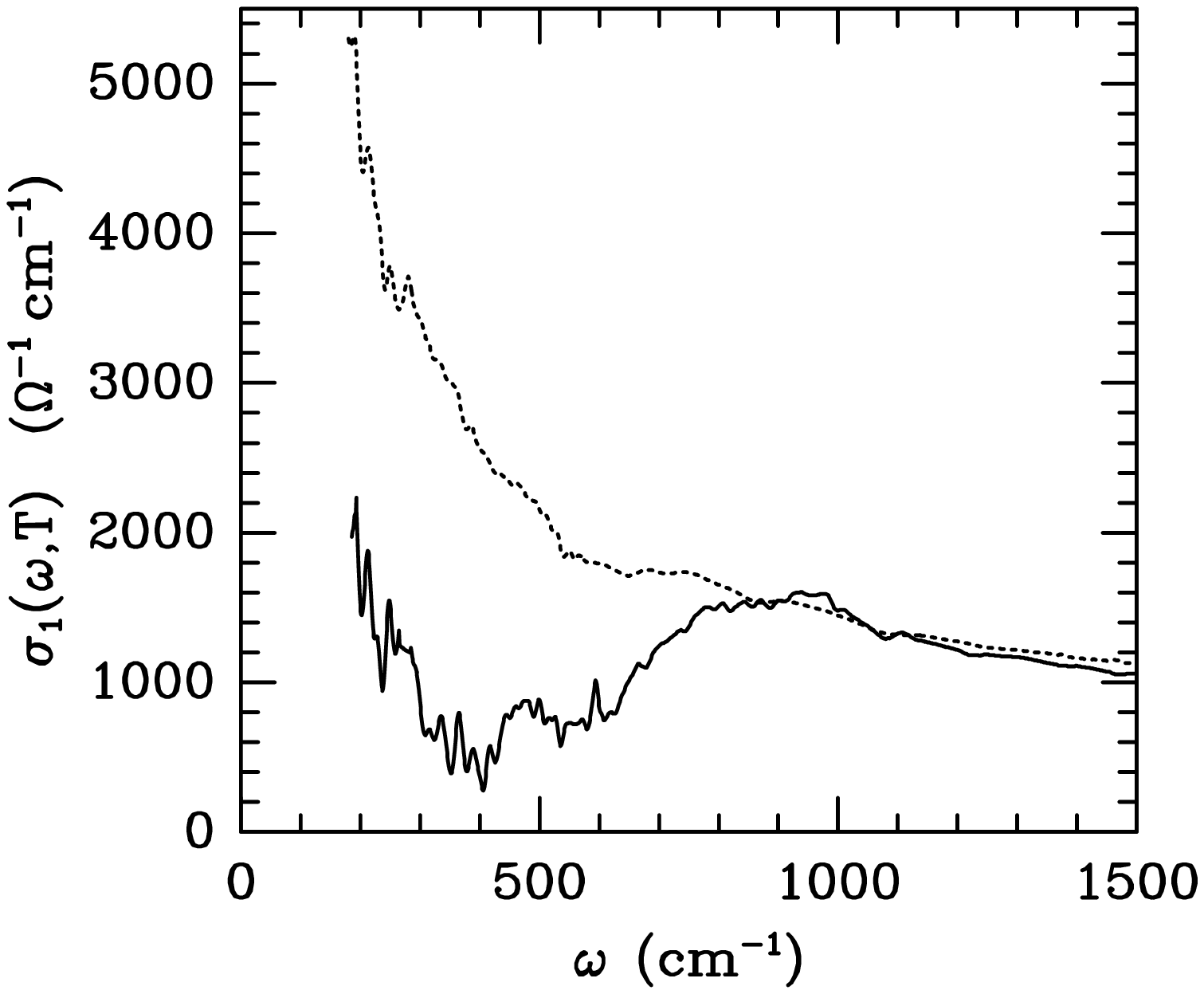}
\caption{Experimental results for the real part of the $a$-axis
conductivity in the normal, $T=100\,$K (dotted line), and
superconducting, $T=20\,$K (solid line), states of an untwinned
YBCO crystal.\protect\cite{Basov1}}
\label{sig1exp}
\end{figure}

The solid curves in Fig.~\ref{sig1exp} show experimental results for
the $a$-axis conductivity $\sigma_1(\omega)$ of an untwinned
YBa$_2$Cu$_3$O$_{6.93}$ crystal.\cite{Basov1} The $a$-axis
conductivity, which avoids the chains, is believed to provide a probe
of the properties of the CuO$_2$ planes.  In the normal state
$\sigma_1(\omega)$
exhibits a much larger spectral weight at high frequencies than would
be present from a Drude form with a frequency independent lifetime.
Various interpretations of this have been suggested.  It has been
proposed\cite{Tanner} that this behavior reflects a two-component
response consisting of a ``free-carrier'' Drude piece and a
mid-infrared (MIR) contribution associated with ``bound-carriers.''
Alternatively, a one-component quasiparticle description in which the
quasiparticle scattering rate $\tau^{-1}=b\max(T,\omega)$ has also
been used to fit the normal state data.\cite{Varma} However, a
consistent fit with constant prefactor $b$ cannot be obtained in
frequency ranges both above and below $\sim 1000\,{\rm
cm}^{-1}$.\cite{Nicoletal}

Likewise, in the superconducting state, various explanations of the
data have been proposed.  In the two-component picture, it is
argued\cite{Tanner,Tim,Kam} that the ``free-carriers'' condense to
form the superfluid while the ``bound-carriers'' remain, giving rise
to the MIR structure which becomes more clearly visible below $T_c$.
In this view the mean free path of the free-electron component is
large compared to $\xi_0$ so that one is in the clean limit in which
it is difficult to see the $2\,\Delta$ onset of $\sigma_1(\omega)$.
Moreover, the
MIR absorption acts to obscure any small onset feature.  An
alternative explanation\cite{Zac} assigned the structure beginning at
500$\,$cm$^{-1}$ to an $s$-wave $2\,\Delta$ onset, giving
$2\,\Delta_0/kT_c \sim 8$.  The problem in this case is to understand
the low-temperature absorption clearly visible below $2\,\Delta$.

Recently, following various experiments\cite{Proceed} which suggest
that the high-temperature cuprates are characterized by a non-$s$-wave
gap, there have been several explanations for the infrared data based
on a gap with nodes.  Putikka and Hirschfeld calculated the ab plane
conductivity for a $d_{x^2-y^2}$ state over
a cylindrical Fermi surface, including impurity scattering with
arbitrary phase shift.\cite{PH} Carbotte {\it et al.}\cite{Carbotte} have
reported $\sigma_1(\omega)$ calculations for a model with a gap
$\Delta_0\cos\theta$.  They include impurity scattering within the
Born approximation and treat the dynamics within a Migdal-Eliashberg
approximation in which a Pb spectrum or alternatively a cut-off
Lorentzian are used as models for the electron-boson spectral density.
Basov {\it et al.}\cite{Basov2} have discussed the infrared conductivity in
YBa$_2$Cu$_3$O$_{6.95}$ crystals which have been ion irradiated in
terms of a possible $d$-wave state and carrier localization produced
by the ion irradiation.  Sauls and co-workers have calculated both the
in-plane and c-axis conductivities
of a $d_{x^2-y^2}$ superconductor including impurity
scattering with arbitrary phase shift but neglecting inelastic
scattering.\cite{Sauls}

Here we propose to extend previous calculations of the microwave
response for a $d_{x^2-y^2}$-wave superconductor to the infrared regime.  Our
basic strategy is the same as in the previous work:\cite{Hirschfeld}
we seek a minimal model consisting of a $d_{x^2-y^2}$ BCS superconductor with
quasiparticles whose inverse lifetimes are given by the sum of elastic
impurity scattering and inelastic spin-fluctuation scattering
$1/\tau\equiv 1/\tau_ {\rm in}+1/\tau_{\rm imp}$.  The inelastic scattering
rate is calculated from a spin-fluctuation exchange interaction which
depends upon the $d_{x^2-y^2}$ gap and was previously found to give a
reasonable fit to the temperature-dependent quasiparticle lifetime
determined by the microwave measurements.\cite{Bonn} For the present
case, the dynamic frequency dependence of the lifetime will be
important.  While a more complete theory is needed, we believe that it
is useful to understand to what extent this minimal theory, which has
previously been used in the microwave regime with considerable
success, can provide a basis for understanding the infrared data.
We find that a number of qualitative
features of the data may be understood within the current framework,
but that some quantitative discrepancies remain which may provide
clues to the physics missing in the minimal model.

In Section II we discuss the model and lay out the calculation of
$\sigma_1(\omega,T)$.  Section III contains the results and
comparisons with data.  Our conclusions are given in Section IV.

\section{Infrared Conductivity}

Within the BCS framework, the real part of the $x$-axis conductivity,
here labeled as $\sigma_1(\omega)$, is given by
\begin{eqnarray}
\sigma_1(\omega) = -{\mathop{\rm Im}\,\Lambda_{xx}(\omega)\over\omega}\ ,
\label{sigma_xx}
\end{eqnarray}
with
\begin{eqnarray}
\mathop{\rm Im}\,\Lambda_{xx}(\omega) = \pi\,e^2
  \int {d^2p\over(2\,\pi)^2}\, d\omega'\, \hbox{Tr}
\left[ \underline A ({\bf p},\omega+\omega')
     \underline A({\bf p},\omega')\right]
     \left[f(\omega+\omega') - f(\omega')\right]
          \left[v_x({\bf p})\right]^2\ .
\label{lambda_xx}
\end{eqnarray}
Here $v_x({\bf p})$ is the electron group velocity along the $x$-direction
and $\underline A({\bf p},\omega)$ is the spectral weight of the Nambu
propagator.
That is, $\underline A({\bf p},\omega) =
-\mathop{\rm Im}\underline{G}({\bf p},\omega)/\pi$, with
\begin{eqnarray}
\underline G({\bf p},\omega) = {\widetilde\omega\underline{\tau}^0 +
\tilde\varepsilon_{{\bf p}}\underline{\tau}^3 +
\widetilde\Delta_{{\bf p}}\underline{\tau}^1 \over
   \widetilde\omega^2 - \tilde\varepsilon^2_{{\bf p}} -
\widetilde\Delta^2_{{\bf p}}}\ ,
\end{eqnarray}
and the $\underline\tau^i$ are Pauli matrices.
The tilde symbols
indicate inclusion of self-energy corrections: \ $\widetilde\omega =
\omega - \sum_0$, $\tilde\varepsilon_{{\bf p}} = \varepsilon_{{\bf p}}
+ \sum_3$, and
$\widetilde\Delta_{{\bf p}} = \Delta_{{\bf p}} + \sum_1$.
In the calculations which follow we use $\varepsilon_{{\bf p}} =
-2\,t(\cos p_x +
\cos p_y) - \mu$ and $\Delta_{{\bf p}} = a\Delta(T)(\cos p_x - \cos p_y)$,
with the parameter $a$ chosen such that the maximum value of the gap
on the Fermi surface is $\Delta(T)$.  For this
tight-binding band the group velocity mentioned above is
given by $v_x({\bf p}) = 2\,t \sin p_x$.

\begin{figure}
\epsfysize=6.5in
\epsfbox{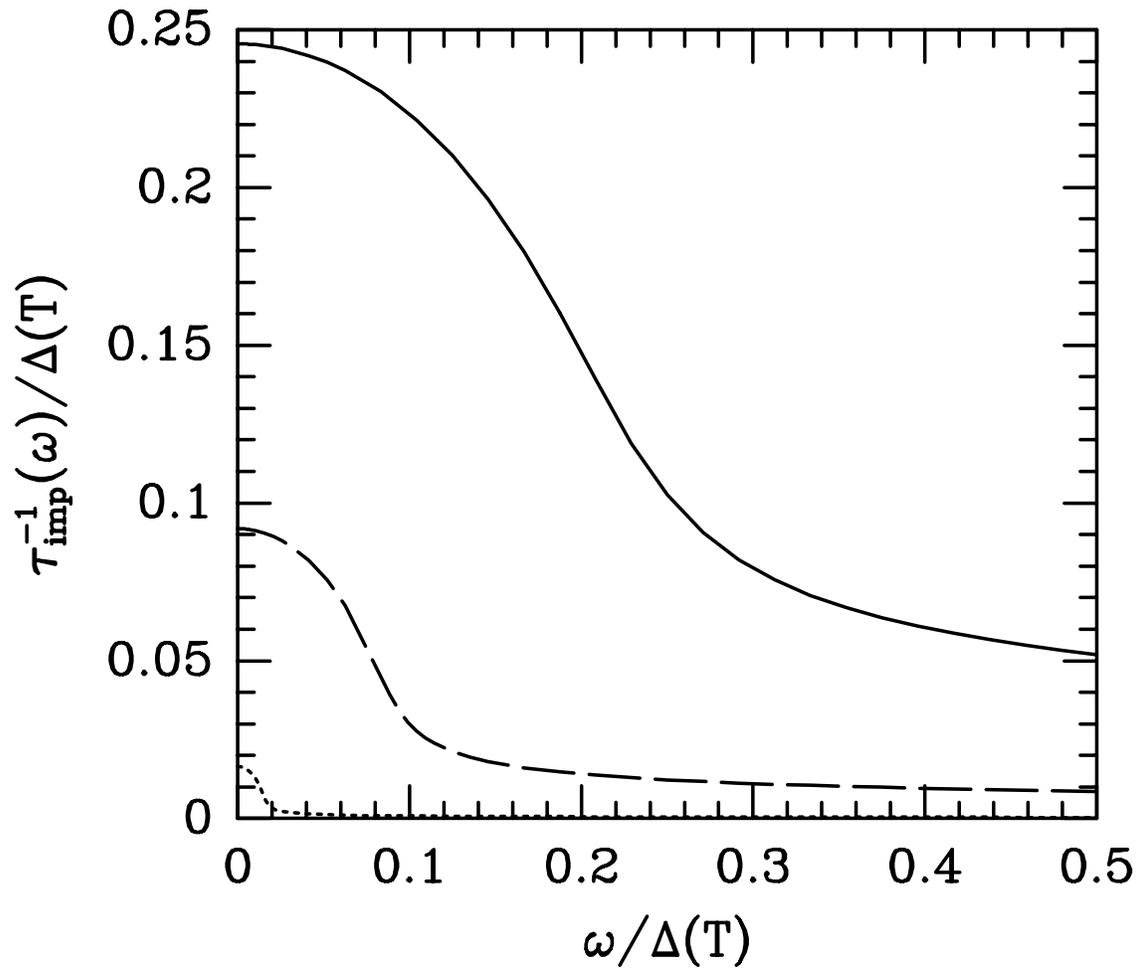}
\caption{Quasiparticle relaxation rate $\tau^{-1}_{\rm imp}$ due to impurity
scattering in the unitary limit, $c=0$. Results are shown for
$\Gamma=0.0008T_c$ (dotted line), $\Gamma=0.018T_c$ (dashed line),
$\Gamma=0.1T_c$ (solid line).}
\label{tauimp}
\end{figure}

The effect of impurity scattering is included by allowing the electron
self-energy to include multiple scattering from a random site potential.
In this case, the self-energy is given by\cite{xx}
\begin{eqnarray}
\Sigma^{\rm imp}_0(\omega) &&= {\Gamma g_0(\omega) \over
c^2 - g^2_0(\omega)}\ ;\nonumber\\ \\
\Sigma^{\rm imp}_3(\omega) &&= {-\Gamma c \over c^2-g^2_0(\omega)}\ .
\nonumber
\end{eqnarray}
Here $\Gamma = n_i/(\pi N_0)$, $c=\cot \delta_0$, and $g_0(\omega) =
1/(\pi N_0) \int d^2p\,\hbox{Tr} [\underline G
({\bf p},\omega)]/(2\,\pi)^2$, where $n_i$ is the impurity concentration,
$N_0$ is the normal phase density of states, and $\delta_0$ is the
scattering phase shift.  The self-energy correction to the gap
function $\Sigma_1$ vanishes for a $d$-wave gap, and in the unitary
limit, $c=0$, only the $\Sigma_0$ contribution remains.  In this case,
the quasiparticle relaxation rate due to the impurity scattering is
\begin{eqnarray}
\tau^{-1}_{\rm imp}(\omega) = -2\,\mathop{\rm Im}\Sigma^{\rm imp}_0(\omega)\ .
\end{eqnarray}
Results for $\tau^{-1}_{\rm imp}$ versus $\omega$ for several values
of $\Gamma$ are shown in Fig.~\ref{tauimp}. The two smaller values of
$\Gamma$ were used in Ref.~\onlinecite{Hirschfeld} to fit penetration
depth measurements and are used again here to allow comparison with
those previous results.

In order to take into account the dynamic spin-fluctuation contributions
to the quasiparticle lifetime, we include, in addition to the impurity
scattering, the imaginary part of the self-energy that arises from
spin-fluctuation exchange
\begin{small}
\begin{eqnarray}
\mathop{\rm Im}\,\underline\Sigma^{\rm sf}({\bf p},\omega) =
  &&\int{d^2p'\over(2 \pi)^2}\>{1 \over 2 [f(\omega)-1]}\cr
  &&\times
  \Biggl\{ \mathop{\rm Im}\,V({\bf p}-{\bf p}',\omega-E_{{\bf p}'}) \,
  [n(\omega-E_{{\bf p}'})+1][1-f(E_{{\bf p}'})] \,
  \left(\underline\tau^0 + {\varepsilon_{{\bf p}'}
  \over E_{{\bf p}'}}\,\underline\tau^3
  + {\Delta_{{\bf p}'} \over E_{{\bf p}'}}\,\underline\tau^1\right)\cr
  &&\quad{}+ \mathop{\rm Im}\,V({\bf p}-{\bf p}',\omega+E_{{\bf p}'})
  \, [n(\omega+E_{{\bf p}'})+1]\,f(E_{{\bf p}'})
   \,
  \left(\underline\tau^0 - {\varepsilon_{{\bf p}'}
  \over E_{{\bf p}'}}\,\underline\tau^3
  - {\Delta_{{\bf p}'} \over E_{{\bf p}'}}\,\underline\tau^1\right)
  \Biggr\}\ .
\label{im_sigma_sf}
\end{eqnarray}
\end{small}
Here $n$ and $f$ are the usual Bose and Fermi functions, and
\begin{eqnarray}
V({\bf q},\omega) =
{3\over2}\, {\overline U^2 \chi_{_{\scriptstyle0}}({\bf q},\omega) \over 1 -
\overline U \chi_{_{\scriptstyle0}}({\bf q},\omega)}\ ,
\end{eqnarray}
with
\begin{eqnarray}
\chi_{_{\scriptstyle0}}({\bf q},\omega) = \int {d^2p\over(2\pi)^2}
&&\Biggl\{ {1\over2}
        \left[ 1 + {\varepsilon_{{\bf p}+{\bf q}}\varepsilon_{{\bf p}}
+ \Delta_{{\bf p}+{\bf q}}\Delta_{{\bf p}}
\over E_{{\bf p}+{\bf q}}E_{{\bf p}}} \right]
     {f(E_{{\bf p}+{\bf q}}) - f(E_{{\bf p}}) \over \omega -
     (E_{{\bf p}+{\bf q}} - E_{{\bf p}})+i\delta}\cr
\noalign{\medskip}
&&\quad {} + {1\over4} \left[
1-{\varepsilon_{{\bf p}+{\bf q}}\varepsilon_{{\bf p}}+\Delta_{{\bf
p}+{\bf q}}\Delta_{{\bf p}}\over
E_{{\bf p}+{\bf q}}E_{{\bf p}}}\right]
   {1-f(E_{{\bf p}+{\bf q}})-f(E_{{\bf p}}) \over \omega_m +
   (E_{{\bf p}+{\bf q}}+E_{{\bf p}}) + i\delta}\cr
\noalign{\medskip}
&&\quad {}+ {1\over4} \left[1 - {\varepsilon_{{\bf p}+{\bf q}}
\varepsilon_{{\bf p}}+\Delta_{{\bf p}+{\bf q}}\Delta_{{\bf p}}
\over E_{{\bf p}+{\bf q}}E_{{\bf p}}}\right]
   {f(E_{{\bf p}+{\bf q}}) + f(E_{{\bf p}}) - 1 \over
           \omega - (E_{{\bf p}+{\bf q}}+E_{{\bf p}}) + i\delta} \Biggr\}\ ,
\label{chi_0}
\end{eqnarray}
and $\overline U$ a phenomenological interaction parameter.

{}From Eq.~(\ref{im_sigma_sf}) it follows that
The inelastic spin-fluctuation induced lifetime of a quasiparticle of
energy $\omega$ and momentum ${\bf p}$ in a superconductor at temperature $T$
can be written as
\begin{small}
\begin{eqnarray}
\tau^{-1}_{\rm in}({\bf p},\omega) =
  \int {d^2p'\over(2\,\pi)^2} \left[ {1\over 1-f(\omega)} \right]
   \Biggl\{&& \int^{\omega-|\Delta_{{\bf p}'}|}_0 d\nu\, \mathop{\rm Im}\,
V({\bf p}-{\bf p}',\nu)\delta\left(\omega-\nu-E_{{\bf p}'}\right)\cr
      &&\qquad\qquad {}\times\left[ 1 + {\Delta_{\bf p}\Delta_{{\bf p}'}
                                + \varepsilon_{\bf p}\varepsilon_{{\bf p}'}
          \over \omega(\omega-\nu)} \right] [n(\nu)+1][1-f(\omega-\nu)] \cr
   &&{}+ \int^0_{\omega+|\Delta_{{\bf p}'}|} d\nu\,\mathop{\rm Im}\,
          V({\bf p}-{\bf p}',\nu)\delta\left(\nu-\omega-E_{{\bf p}'}\right)\cr
     &&\qquad\qquad {}\times\left[ 1 - {\Delta_{\bf p}\Delta_{{\bf p}'}
                                + \varepsilon_{\bf p}\varepsilon_{{\bf p}'}
           \over\omega(\nu-\omega)}\right] [n(\nu)+1][f(\nu-\omega)]\cr
   &&{}+ \int^\infty_0 d\nu\,\mathop{\rm Im}\,V({\bf p}-{\bf p}',\nu)
             \delta\left(\omega+\nu-E_{{\bf p}'}\right)\cr
        &&\qquad\qquad {}\times\left[ 1 + {\Delta_{\bf p}\Delta_{{\bf p}'}
                                + \varepsilon_{\bf p}\varepsilon_{{\bf p}'}
                \over \omega(\omega+\nu)}\right]
         n(\nu) [1-f(\omega+\nu)] \Biggr\}\ .
\label{tau_in}
\end{eqnarray}
\end{small}
The first and third terms represent scattering processes associated
with the emission and absorption of spin fluctuations, while the
second term arises from the recombination of two quasiparticles to
form a pair.  Above $T_c$, $\Delta_{\bf p}$ goes to zero and
Eq.~(\ref{tau_in}) reduces to the usual normal state expression.

\begin{figure}
\epsfysize=6.5in
\epsfbox{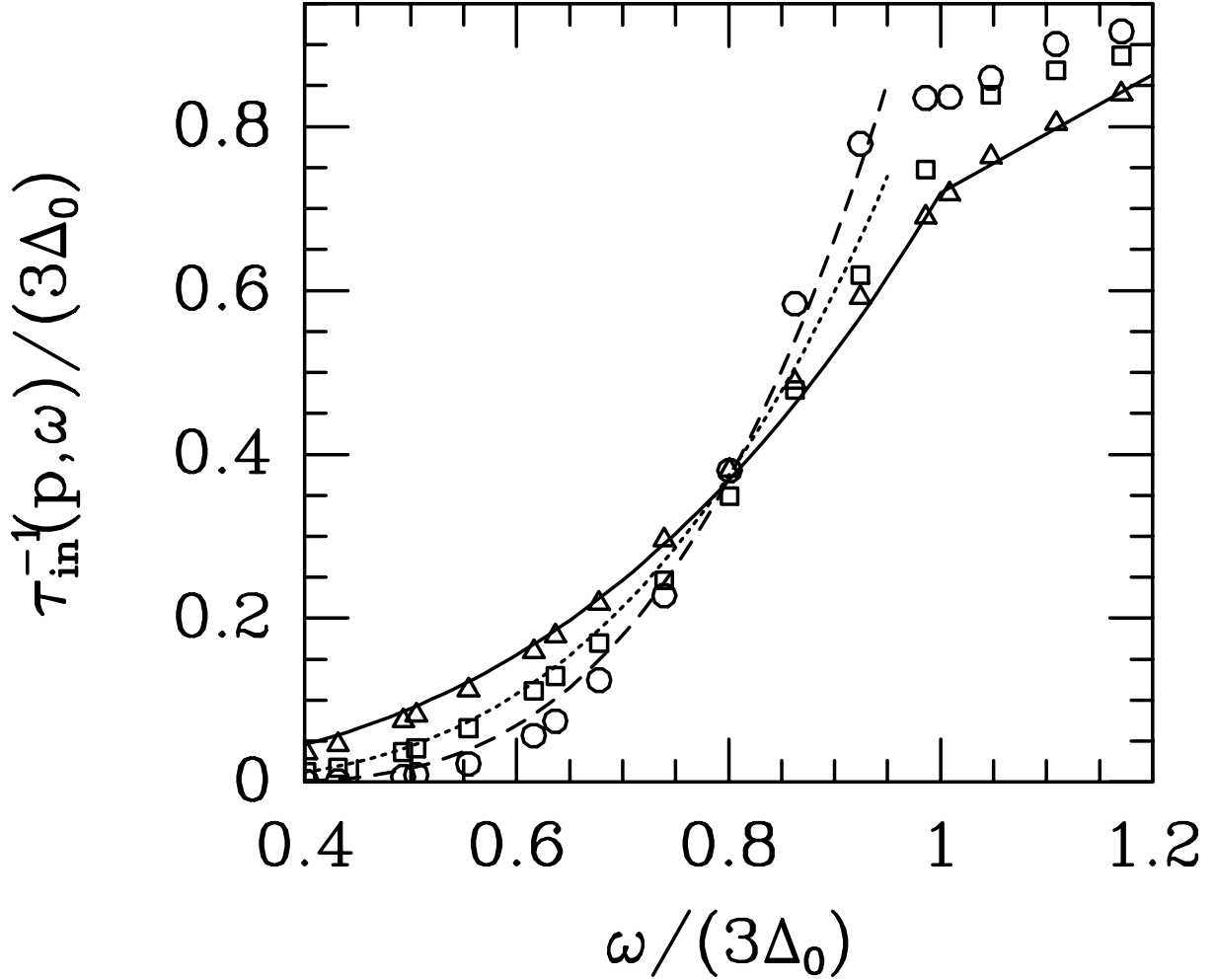}
\caption{Quasiparticle relaxation rate $\tau^{-1}_{\rm in}({\bf p},\omega)$
due to
spin-fluctuation scattering. Results are shown as a function of
frequency for $T=0.1T_c$. The different symbols indicate values
calculated for ${\bf p}$ at three different points along the Fermi surface:
a gap node (triangles), a gap antinode (circles), and a point halfway
between the node and the antinode (squares). The solid line shows an
interpolation which varies as $\omega$ for $\omega>3\Delta_0$ and as
$\omega^3$ for $\omega < 3\Delta_0$. The dashed lines vary as
$[\omega-\Delta_{\bf p}]^3$ with ${\bf p}$, as above, at an antinode (long
dashes) and an intermediate point (short dashes). }
\label{nodevsantinode}
\end{figure}

Using parameter values for the two-dimensional Hubbard model
($\langle n\rangle=0.85$, $\overline U=2$, $T_c = 0.2\,t$, $\Delta_0 = 3\,T_c$)
which were previously used to model NMR relaxation time
data,\cite{Bulut2} calculations of the temperature dependence of the
inelastic scattering lifetime $\tau^{-1}_{\rm in}({\bf p},\omega=T)$ were
previously reported.\cite{Quin} Here we examine the frequency
dependence of $\tau^{-1}_{\rm in}({\bf p},\omega)$. Figure
\ref{nodevsantinode} shows $\tau^{-1}_{\rm in}$ versus $\omega$ for
three different Fermi surface momenta.\cite{vep-note} At a node,
$\tau^{-1}_{\rm in}({\bf p},\omega)\sim\omega^3$ at low energies crossing
over to an approximate linear variation when
$\omega\gtrsim 3\,\Delta(T)$.  This is similar to the temperature
dependence of $\tau^{-1}_{\rm in}({\bf p},\omega=T)$ previously reported.
At low energies, the usual quasiparticle-quasiparticle Coulomb
scattering would vary as $\omega^2$.  Here the extra power of $\omega$
reflects the low-energy variation of the $d_{x^2-y^2}$-wave density of
states.  For energies larger than several $\Delta(T)$, $\tau^{-1}_{\rm
in}({\bf p},\omega)$ is expected to approach its normal state variation
which is approximately linear for strong spin-fluctuation scattering
and a nearly nested Fermi surface. Also apparent in
Fig.~\ref{nodevsantinode} is that for ${\bf p}$ away from a gap node
$\tau^{-1}_{\rm in}({\bf p},\omega)$ varies approximately as
$(\omega-\Delta_{\bf p})^3$ at low energy, crossing over to linear
variation at higher energy.

The effects of both the elastic and inelastic quasiparticle lifetimes
are included in the infrared conductivity calculations which follow by
adding the electron self-energies due to elastic and inelastic processes
\begin{eqnarray}
\underline\Sigma({\bf p},\omega) = \underline\Sigma^{\rm sf}({\bf p},\omega) +
\underline\Sigma^{\rm imp}(\omega)\ .
\end{eqnarray}
This total self-energy is then used in the evaluation of
Eq.~(\ref{lambda_xx}).

\begin{figure}
\epsfysize=6.5in
\epsfbox{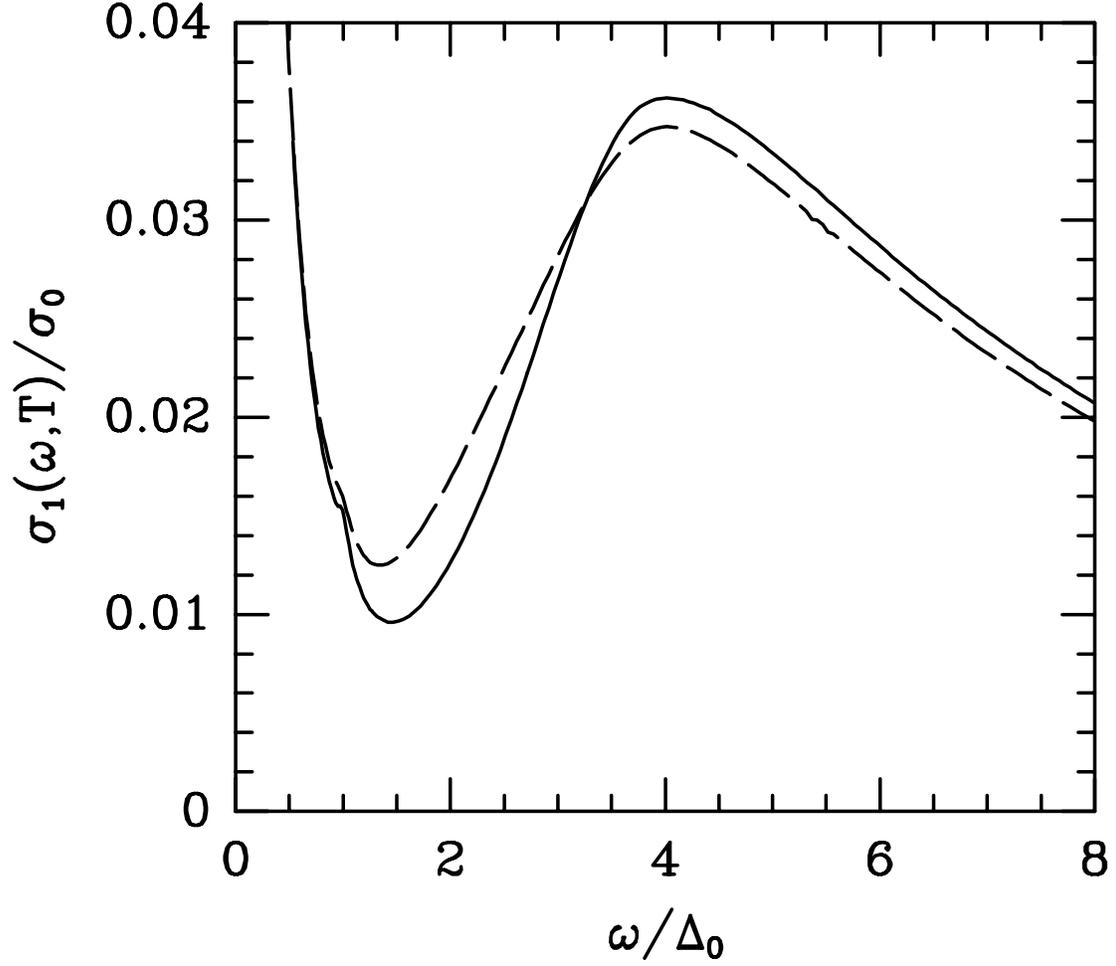}
\caption{Real part of the in-plane conductivity in the superconducting
state for $T=0$ and $\Gamma=0.018T_c$. The two curves show
results for two different choices for the self energy due to
spin-fluctuation scattering. The solid curve incorporates the full
momentum dependence reflected in
Fig.~\protect\ref{nodevsantinode}. The dashed curve replaces
$\Sigma^{\rm sf}({\bf p},\omega)$ everywhere by its value at the gap node on
the Fermi surface.}
\label{sig1vstauin}
\end{figure}

To make further progress we make certain simplifying approximations
for calculational simplicity.
$\mathop{\rm Re}\,\underline\Sigma^{\rm sf}({\bf p},\omega)$
is not included in results presented; test calculations show that it
varies slowly enough with frequency, temperature and momentum to be
absorbed into an effective mass renormalization.
$\mathop{\rm Re}\,\underline\Sigma^{\rm imp}(\omega)$
is retained, however, to allow
for direct comparison with Refs.~\onlinecite{PH} and
\onlinecite{Sauls} at low frequencies. It remains to evaluate
$\mathop{\rm Im}\,\underline\Sigma^{\rm sf}({\bf p},\omega)$, as given in
Eq.~(\ref{im_sigma_sf}). As in Ref.~\onlinecite{Quin}, the momentum
sums in Eqs.~(\ref{im_sigma_sf}) and (\ref{chi_0}) are performed by
evaluating the integrands on a lattice of points covering the
Brillouin zone and summing the results. (For a more detailed
discussion see Ref.~\onlinecite{Quinlanthesis}.) Since the momentum
sum in Eq.~(\ref{chi_0}) is nested inside that of
Eq.~(\ref{im_sigma_sf}), evaluation of
$\mathop{\rm Im}\,\underline\Sigma^{\rm sf}({\bf p},\omega)$
is computationally rather
costly. Performing this evaluation for each point of the momentum
lattice sums required to evaluate Eq.~(\ref{lambda_xx}) would be
prohibitively costly since a separate momentum sum is required for
each value of $\omega'$ needed to numerically evaluate the frequency
integral contained therein. It is therefore a practical necessity to
replace $\mathop{\rm Im}\,\underline\Sigma^{\rm sf}({\bf p},\omega)$ in
Eq.~(\ref{lambda_xx}) by an appropriate interpolation which captures
the most important features of the momentum frequency and temperature
dependence of $\mathop{\rm Im}\,\underline\Sigma^{\rm sf}({\bf p},\omega)$.
As discussed
above, the frequency and momentum dependent of $\tau^{-1}_{\rm
in}({\bf p},\omega)$ (at a temperature well below $T_c$) is reasonably well
fit with an interpolation which varies as $(\omega-\Delta_{\bf p})^3$. One
may ask whether the momentum dependence of this quantity has any
influence on the infrared conductivity. Figure \ref{sig1vstauin} shows
two curves which represent the conductivity calculated with
\begin{eqnarray}
\underline\Sigma^{\rm sf}({\bf p},\omega) =
  -i\left[2\tau^*_{\rm in}({\bf p},\omega)\right]^{-1}\underline\tau_0
\end{eqnarray}
for two different choices of the interpolation function
$\left[\tau^*_{\rm in}({\bf p},\omega)\right]^{-1}$.  One choice is a
$(\omega-\Delta(\theta))^3$ interpolation.  Another choice is to
replace $\underline\Sigma^{\rm sf}({\bf p},\omega)$ by its value at the gap
node on the Fermi surface. This leads to a momentum-independent
interpolation which varies as $\omega^3$ for $\omega<3\,\Delta_0$ and
$\omega$ for $\omega>3\,\Delta_0$.  As may be seen, the two
interpolations produce nearly identical results.  For simplicity of
interpretation then, a momentum-independent interpolation is used in
all of the $\sigma_1(\omega)$ evaluations which follow.

\begin{figure}
\epsfysize=6.5in
\epsfbox{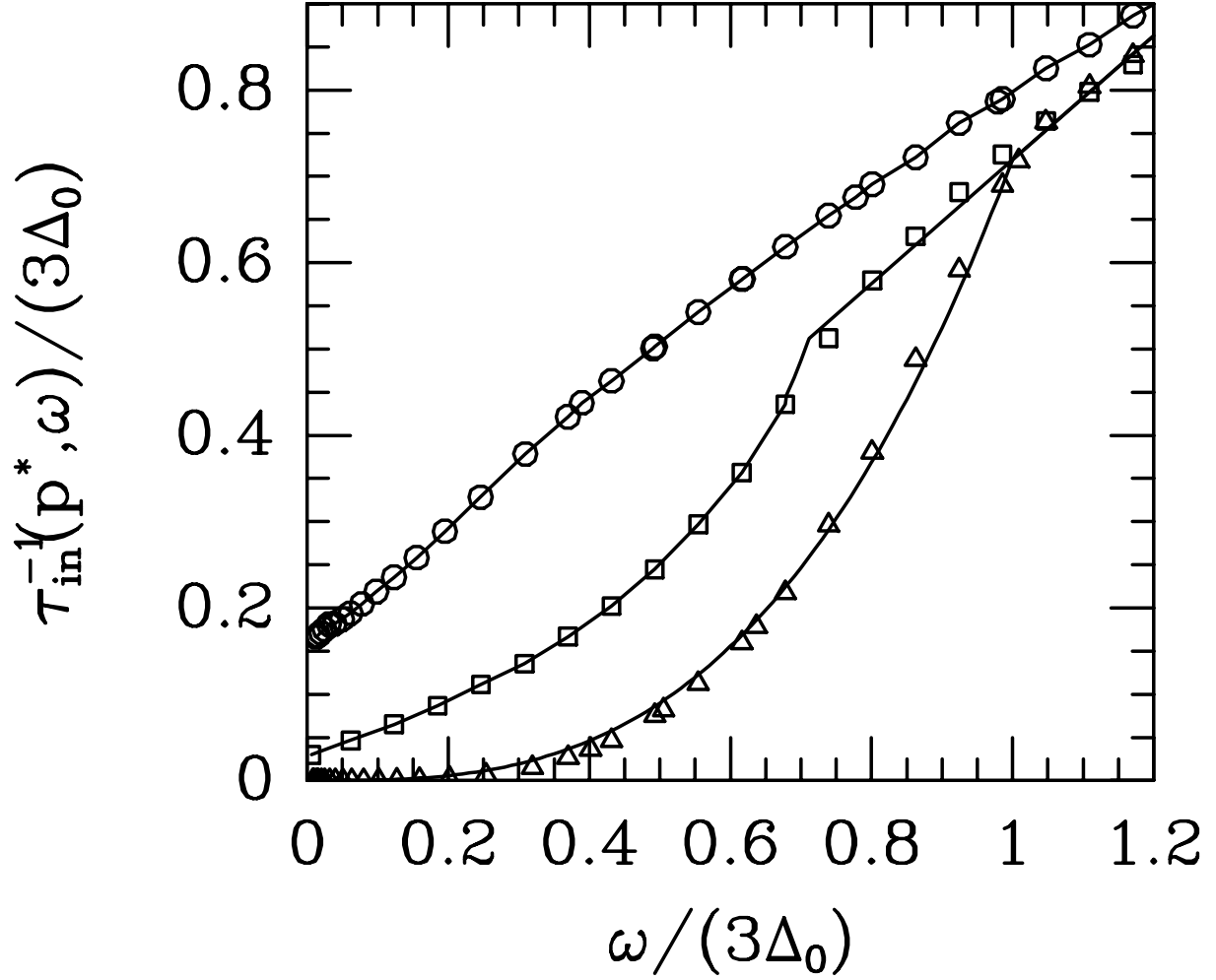}
\caption{Quasiparticle relaxation rate $\tau^{-1}_{\rm in}$ due to
spin-fluctuation scattering. Results are shown as a function of
frequency for $T=T_c$ (circles), $T=0.8T_c$ (squares), and
$T=0.1T_c$ (triangles). The solid lines show interpolations used at
each temperature for the evaluation of $\sigma_1(\omega)$.}
\label{tausf-temp}
\end{figure}

Of course, in order to calculate $\sigma_1(\omega)$ at other than the
lowest temperatures, we must incorporate temperature dependence in the
$\left(\tau^*_{\rm in}\right)^{-1}$ interpolation.  Figure
\ref{tausf-temp} shows $\tau^{-1}_{\rm in}({\bf p}^*,\omega)$, where
${\bf p}^*$ is the Fermi surface wavevector at a gap node, calculated for $T$
equal to $T_c$, $0.8T_c$, and $0.1T_c$.  The low temperature curve
shows the cubic-to-linear crossover at $3\,\Delta_0$ discussed
above. The $T=0.08T_c$ curve also shows a crossover at
$\omega=3\,\Delta(T)$, but varies more slowly than $\omega^3$ at low
temperatures.  The solid curves in Fig.~\ref{tausf-temp} show the
interpolations used at the respective temperatures in the calculations
of $\sigma_1(\omega)$.

Before giving our results, we note that vertex corrections to the
conductivity have been neglected.  While this is justified in
the impurity-dominated regime, where $s$-wave impurity vertex
corrections to current-current correlations functions vanish
identically for singlet states, it represents a further approximation
in the region where the inelastic spin fluctuations provide the
dominant scattering.  We ignore these corrections in what follows in
order to get a qualitative picture of the conductivity in the minimal
model.

\section{Results}

\begin{figure}
\epsfysize=6.5in
\epsfbox{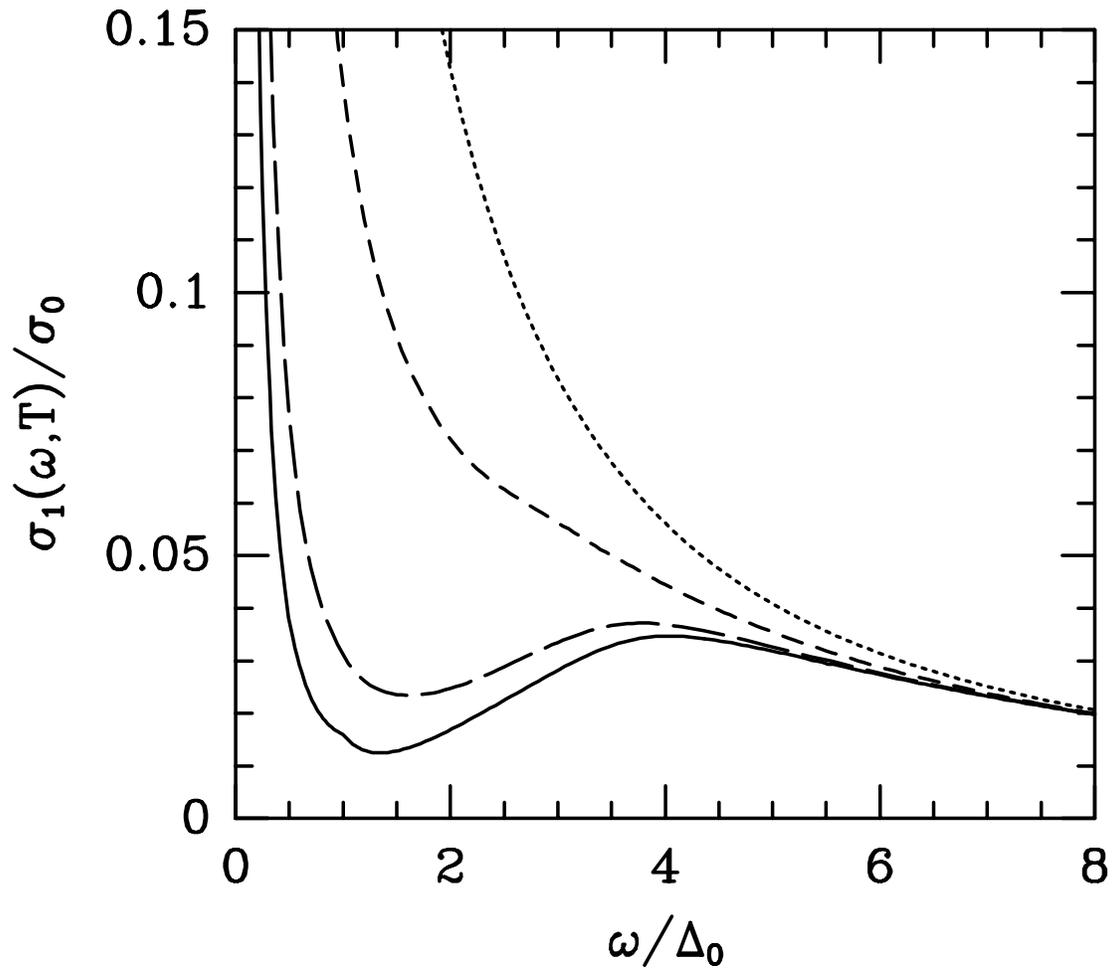}
\caption{Real part of the in-plane conductivity for $\Gamma=0.018T_c$
at several different temperatures. Results are shown for $T=T_c$ (dotted
line), $T=0.8T_c$ (short dashes), $T=0.5T_c$ (long dashes), and
$T=0$ (solid line).}
\label{sig1vstemp}
\end{figure}

\begin{figure}
\epsfysize=6.5in
\epsfbox{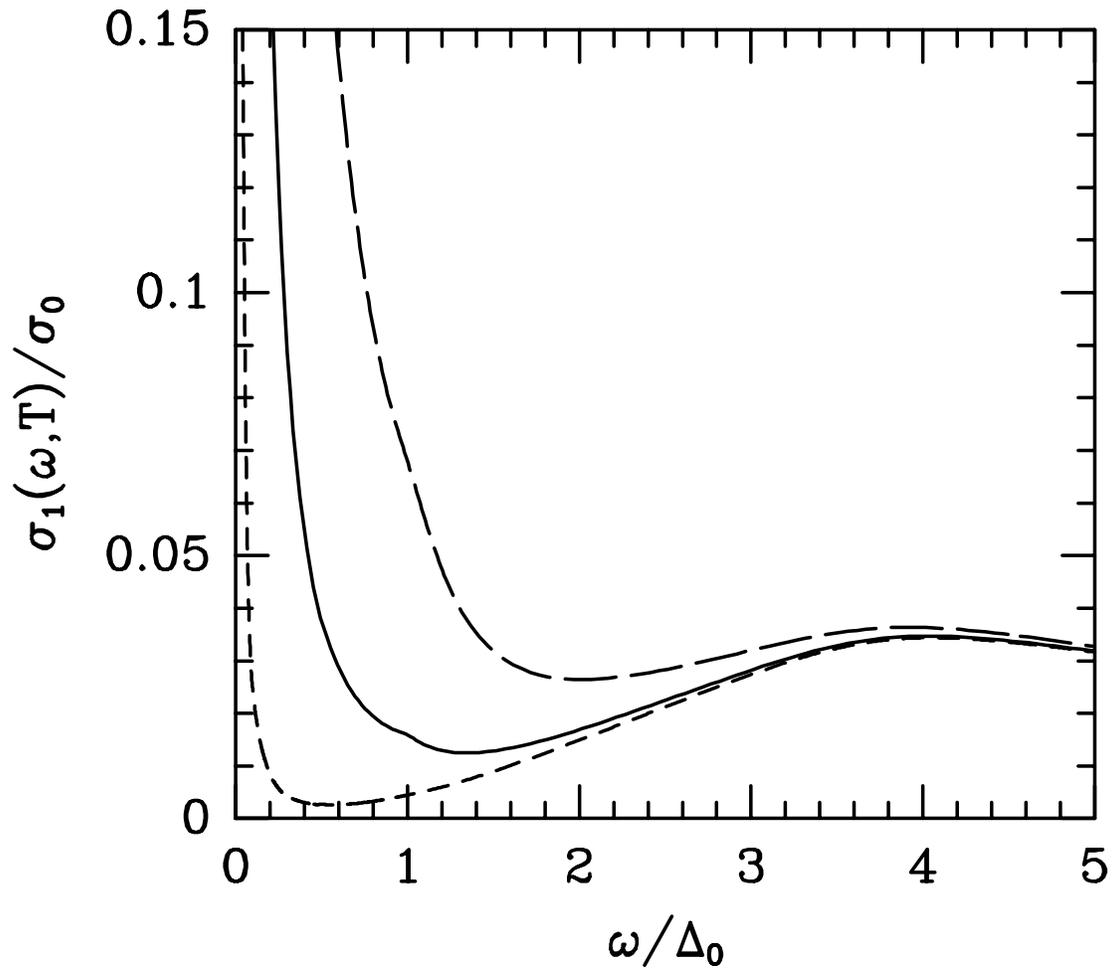}
\caption{Real part of the in-plane conductivity in the
superconducting state ($T=0$). Results are shown for
$\Gamma=0.0008T_c$ (short dashes), $\Gamma=0.018T_c$ (solid line),
and $\Gamma=0.1T_c$ (long dashes)}
\label{sig1vsgamma}
\end{figure}

\begin{figure}
\epsfysize=6.5in
\epsfbox{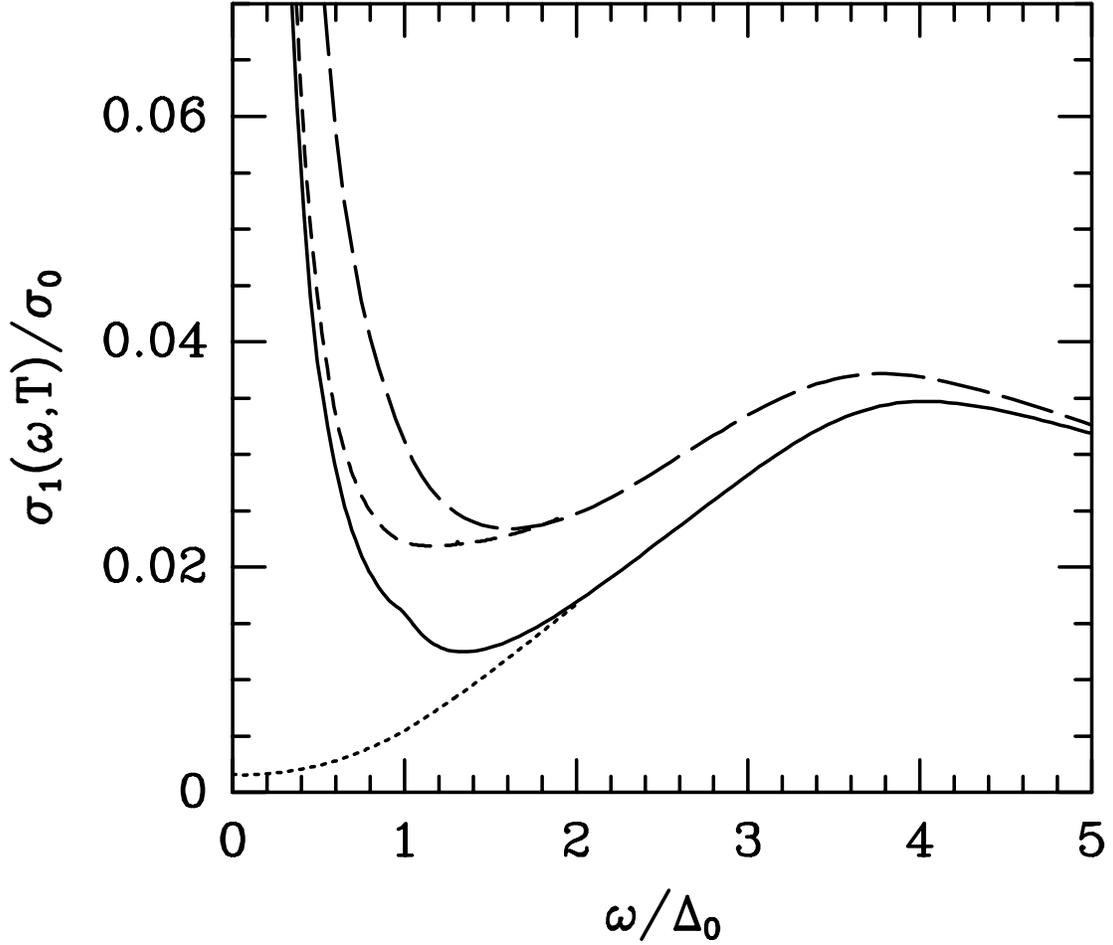}
\caption{Real part of the in-plane conductivity in the superconducting
state for impurity scattering in the Born, $c \gg 1$ (dashed line), and
unitary, $c=0$ (solid line), limits. Curves are shown for $T=0$
(dotted line is the Born limit, solid line is the unitary limit) and
for $T=0.5T_c$ (short dashes show the Born limit, long dashes show the
unitary limit) All curves are calculated using $\Gamma/(c^2+1)=0.018T_c$.}
\label{sig1vsc}
\end{figure}

Our basic results for $\sigma_1(\omega,T)$ are summarized in
Figs.~\ref{sig1vstemp}--\ref{sig1vsc}.  Here $\sigma_1(\omega,T)$,
normalized to $\sigma_0=(ne^2/m)/(2T_c)\approx\sigma_1(0,T_c)$, is
plotted versus $\omega$
for various temperatures and elastic impurity scattering parameters.
A $2\,\Delta_0/kT_c$ ratio of 6 was chosen for these plots.  Figure
\ref{sig1vstemp} shows $\sigma_1(\omega,T)$ for various temperatures
for a unitary impurity scattering strength $\Gamma=0.018\,T_c$.  In
the normal state at $T=T_c$, the inelastic scattering strength is
dominant, leading to an enhancement of the spectral weight over the
simple Drude $\omega^{-2}$ behavior.  As the temperature decreases
below $T_c$ and the $d_{x^2-y^2}$ gap opens, spectral weight is transferred
into the $\omega=0$ superfluid delta-function.  A narrow ``residual
Drude'' response due to scattering by nodal quasiparticles remains,
becoming increasingly narrow and high as $T/T_c$ decreases until the
quasiparticle lifetime is limited by elastic impurity scattering.
Figure \ref{sig1vsgamma} illustrates the effect of increasing the
elastic scattering rate $\Gamma$ which broadens this Drude-like
response.  If the impurity scattering rate in the normal state is
significantly smaller than $\Delta_0$, this feature may be shown to
persist crudely out to a frequency $\omega^*$ determined by the
crossover of the elastic and the inelastic scattering rates, and
corresponding roughly to the position of the low temperature
conductivity minimum in Figs.~\ref{sig1vstemp} and
\ref{sig1vsgamma}. In the clean limit we find $\omega^*\sim (\Gamma
T_c^3)^{1/4}$ if impurity scattering is resonant, and $\omega^*\sim
(\Gamma_N T_c)^{1/2}$ in the weak scattering limit $c\gg 1$ (where
$\Gamma_N=\Gamma/[c^2+1]$). At $T=0$ the weight in the residual Drude
feature is largest in the unitary limit, but increases considerably
more rapidly at finite temperatures for weak scattering, as seen in
Fig.~\ref{sig1vsc}. This may be understood in terms of the slower
$T\rightarrow T^2$ crossover in the penetration depth in this
case,\cite{Arberg,felds} reflected in the temperature dependent part
of the condensate depletion. We also note that the {\it effective}
(experimentally observable) $\omega\rightarrow 0$ conductivity in the
weak scattering limit is of order $\sigma_{\rm imp} \equiv
ne^2/2m\Gamma_N$, much larger than the ``universal''\cite{PALee}
resonant d-wave result $\sigma_{00}\simeq ne^2/m\pi\Delta_0$ in the
clean limit.

In the resonant impurity scattering limit $c=0$, a ``shoulder'' arises
at $\omega=\Delta_0$ in the conductivity due to scattering of
quasiparticles from the scattering resonance in the density of states
at $\omega=0$ to the peak at the gap ``edge.''\cite{pwave} Although
this might in principle be used to identify the existence of strong
elastic scattering, it is smeared by finite temperature effects and
may thus be difficult to observe. It is seen in the $T=0$ conductivity
displayed in Fig.~\ref{sig1vstemp} as a kink around $\Delta_0$.
Note there is no particular structure in the d-wave
conductivity at $\omega=2\,\Delta_0$.

\begin{figure}
\epsfysize=6.5in
\epsfbox{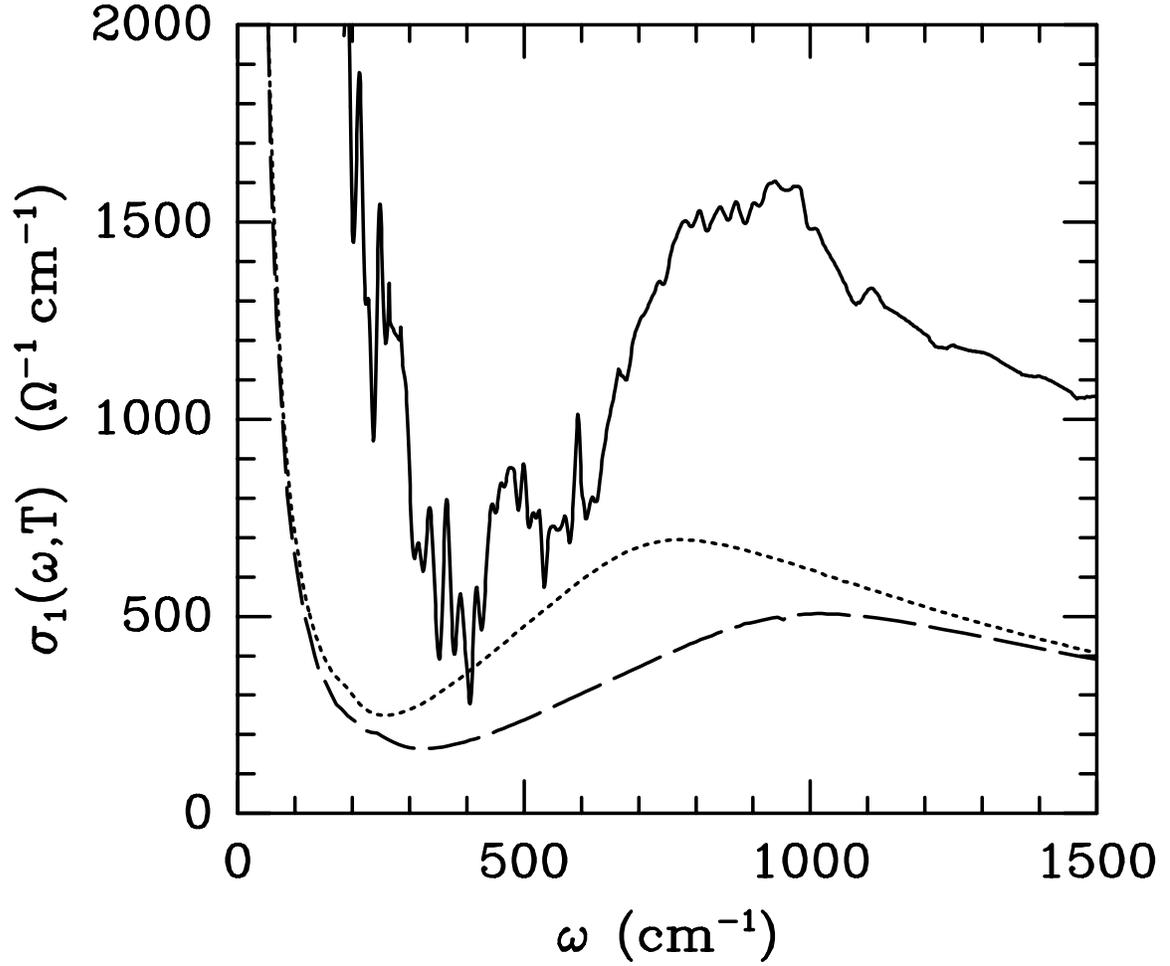}
\caption{Real part of the in-plane conductivity in the superconducting
state calculated for two different choices for the superconducting gap
ratio. Results are shown for $\Delta(0)=3T_c$ (dotted line) and
$\Delta(0)=4T_c$ (dashed line). The solid line shows the experimental
result for $a$-axis conductivity in the superconducting state,
$T=20\,$K, of an untwinned YBCO crystal.\protect\cite{Basov1}}
\label{sig1vsdelta}
\end{figure}

In the limit $\tau^{-1}_{\rm imp}\to0$, the onset at $2\,\Delta_0$ in
an $s$-wave superconductor is suppressed by the inability to conserve
momentum and the onset appears at $4\,\Delta_0$, where inelastic
scattering leads to a four quasiparticle final
state.\cite{Orensteinetal} For the $d_{x^2-y^2}$ case, this effect is
broadened by the gapless nature of the state; however, a broad peak at
$4\,\Delta_0$, with $\Delta_0$ the maximum of the $d_{x^2-y^2}$ gap, is
clearly visible above the minimum conductivity at $\sim\omega^*$. The
peak shifts downward and becomes larger as the quasiparticle
relaxation rate becomes comparable to $\Delta_0$, but is near
$4\,\Delta_0$ in the clean limit, which obtains at these frequencies
if Hubbard parameters consistent with normal state data are chosen.
Figure \ref{sig1vsdelta} shows how this broad peak shifts with
$2\,\Delta_0/kT_c$.

To summarize, a $d_{x^2-y^2}$ BCS model in which quasiparticle damping due to
impurity scattering and spin-fluctuation scattering is included gives
rise to (1) a low-frequency Drude-like feature whose width and
strength depend upon the impurity concentration, and (2) a midinfrared
maximum at $\omega$ of order $4\,\Delta_0$ whose strength depends upon
$2\,\Delta_0/kT_c$ and the strength of the inelastic scattering.

\section{Conclusions}

\begin{figure}
\epsfysize=6.5in
\epsfbox{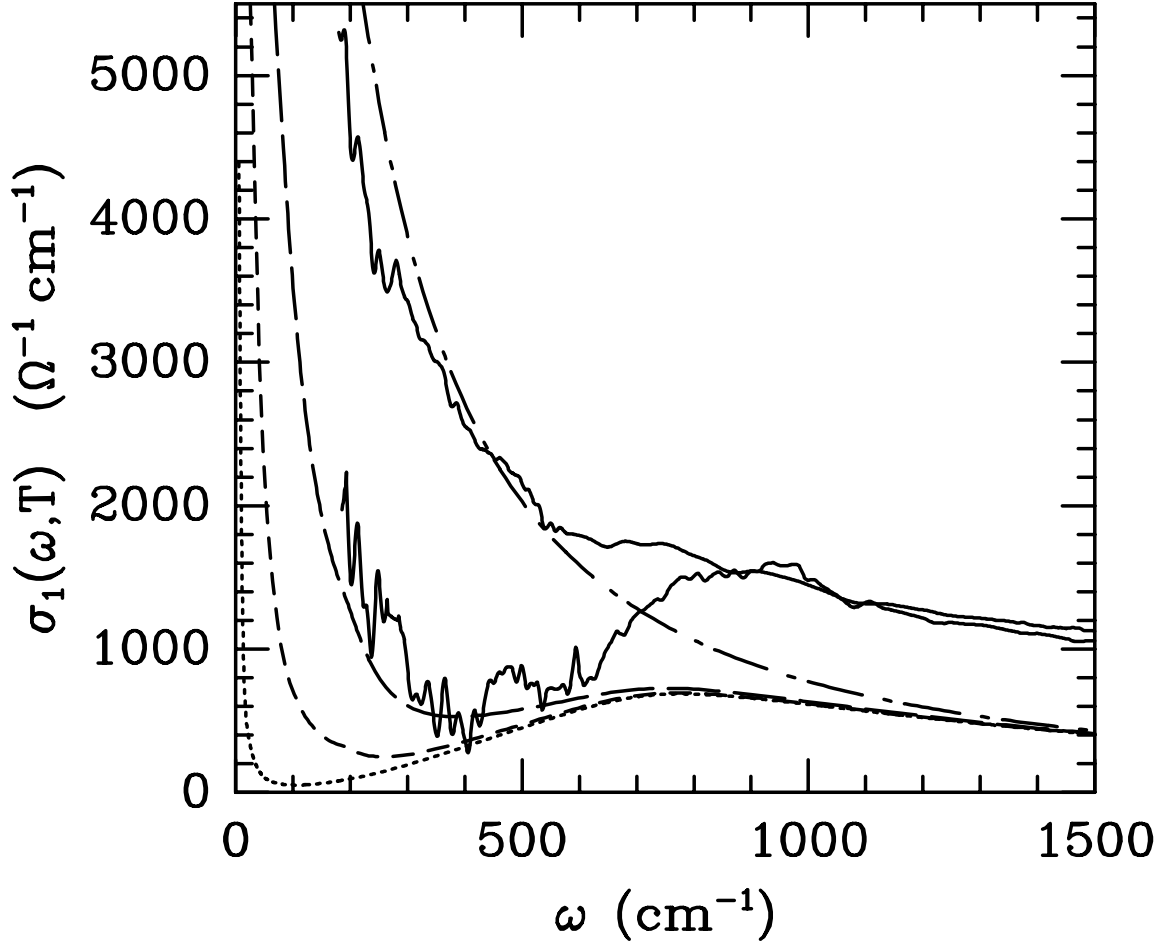}
\caption{Real part of the in-plane conductivity calculated for the
normal, $T=T_c$ and $\Gamma=0.018T_c$ (dot-dashed line), and
superconducting, $T=0.1T_c$,
states compared to experimental results. Calculated superconducting
state results are shown for $\Gamma=0.0008T_c$ (dotted line),
$\Gamma=0.018T_c$ (short dashes), and $\Gamma=0.1T_c$ (long
dashes). The solid lines show experimental results for $a$-axis
conductivity in the normal, $T=100\,$K, and superconducting,
$T=20\,$K, states of an untwinned YBCO crystal.\protect\cite{Basov1}}
\label{sig1vsgammavsexp}
\end{figure}

The dashed lines in Fig.~\ref{sig1vsgammavsexp} show a possible fit of
our model calculation of the $a$-axis conductivity $\sigma_1$ for the
normal and superconducting states of the untwinned YBCO crystal
studied in Ref.~\onlinecite{Ref}. The $a$-axis data was chosen for
comparison to minimize the influence of absorption due to the YBCO
chain layers.  Here we have taken $\sigma_0 =
20,000\,\Omega^{-1}\,$cm$^{-1}$, unitary ($c=0$) impurity scattering
with $\Gamma=0.1\,T_c$, and $2\,\Delta_0/kT_c = 6$.  Note that the
value of $\sigma_0$ is consistent with an effective mass of order 2
and a quasiparticle lifetime at $T_c$ of order $T^{-1}_c$, which is
what we have used.

Figure \ref{sig1vsgammavsexp} demonstrates the qualitative similarity
of the data to the theory, in particular the beginning of the onset of
the residual Drude feature, and a conductivity peak at about
$1000\,{\rm cm}^{-1}$, corresponding roughly to 4$\,\Delta_0$. There
are, however, several apparent discrepancies. Firstly, our model fails
to give sufficient spectral weight at high frequencies.  We believe
that this represents a failure to adequately describe the normal state
at these higher energies, equivalent to the discrepancies encountered
when comparing the Marginal Fermi Liquid model to
experiment.\cite{Nicoletal} It may be, as recent numerical
calculations suggest,\cite{Dagotto,Jarrell,Bulut} that this extra
weight arises from excitations from the lower Hubbard band to the
narrow quasiparticle band which forms at the upper edge of the lower
Hubbard band when the system is doped away from half-filling.  Such
calculations also show a distinct peak at roughly $J\simeq 1000\,{\rm
cm}^{-1}$ in the normal state conductivity. However,
a complete explanation along these lines should provide an
understanding not only for the MIR peak in the normal state of
YBa$_2$Cu$_3$O$_{6.6}$ and La$_{2-x}$Ba$_x$CuO$_4$, but also for its
absence in YBa$_2$Cu$_3$O$_7$ and Bi$_2$Sr$_2$CaCu$_2$O$_8$. It has
been speculated that the latter,
``spin gap'' materials consist in their normal state of preformed
pairs which form a superconducting condensate below
$T_c$.\cite{preformed} Such pair correlations might then give rise
even for $T>T_c$ to the inelastic MIR peak at $4\,\Delta_0$ found in
the current mean field theory, whereas in the former class of
compounds the coherence and pairing onset temperatures coincide, so a
distinct MIR peak is seen only for $T<T_c$. Finally, we note that we
are not able to rule out the possibility of a separate MIR
band of electronic excitations which do not participate in
superconductivity.\cite{Tanner}

A second discrepancy arises when we try to compare values of impurity
scattering rates found by fitting optical data as in
Fig.~\ref{sig1vsgammavsexp} with those deduced from fits to the
low-temperature penetration depth.  In the unitary scattering limit,
it was found in Ref.~\onlinecite{Hirschfeld} that values of $\Gamma$
of roughly $10^{-3} T_c$ were characteristic of nominally pure high
quality single crystals. As seen in Fig.~\ref{sig1vsgammavsexp},
low-frequency optical data on similar samples requires a defect
scattering rate a hundred times larger within the same model.  This
might be taken to suggest that our treatment of the impurity t-matrix
is too naive at this point.  Note further that in order to obtain a
consistent picture at both microwave and infrared frequencies, it is
not sufficient to simply work in the Born limit,\cite{Carbotte} since
in general increasing $c$ at fixed normal state scattering rate
decreases the weight in the residual Drude peak (Fig.~\ref{sig1vsc}).

However, it is important to recognize that the upturn in the
$\sigma_1(\omega)$ data, obtained by a Kramers-Kr\"onig transform of
reflectance data, is extremely sensitive to the determination of the
100\% reflectance point, a difficult task since resolution decreases
at the lowest frequencies. Transmittance
experiments,\cite{BSCCOtransmittance} which are not subject to this
difficulty, appear to show a {\it downturn} in $\sigma_1(\omega)$ in
this frequency range. This might be consistent with a true excitation
gap, but would also be consistent with a very narrow residual Drude
feature below the optical resolution limit, since as
$\omega\rightarrow 0$ the conductivity must rise to its DC value of
roughly $2\times 10^4\,\Omega^{-1}\,{\rm cm}^{-1}$. We
therefore believe that the apparent disagreement with experiment at
low frequencies is not to be taken seriously at this time.
Measurements in the far-infrared crossover region will be
extremely useful to settle this point.

A clue to the phase shift due to defect scattering in these materials
may be provided by a recent measurement by Homes {\it et al.}, in which a
shoulder was observed at $300\,{\rm cm}^{-1}$ only at the lowest
measurement temperature (6$\,$K but not at 12$\,$K) in a Ni-doped
sample of optimally doped YBCO.\cite{Homesetal} This small but
identifiable feature is quite close to the frequency $\Delta_0$ where
we expect a shoulder due to unitary limit scattering from the Fermi
level resonance to the BCS gap edge, provided $\Delta_0/T_c\simeq$ 3-4.

Despite the discrepancies outlined above, the qualitative fit of this model
suggests that (1) the superconducting state of YBCO is characterized
by a gap with nodes and (2) the inelastic scattering at energies less
than several $\Delta_0$ is suppressed in the superconducting state.  A
gap with nodes can give rise to the residual Drude weight
observed for $T\ll T_c$, and to an increased
quasiparticle density as the impurity scattering increases due to
pair-breaking processes which are present for a non-$s$-wave gap. This
conclusion is supported by the irradiation experiments of Basov
{\it et al.}\cite{Basov2} and the Ni-doping studies of Homes
{\it et al.},\cite{Homesetal} where the weight in the residual Drude feature
clearly increases with disorder.  The way in which the inelastic
scattering is suppressed at frequencies less than several $\Delta_0$
implies that the dominant dynamic interaction is electronic in nature
so that its spectral weight in this region is reduced when the gap
opens.  The observed structure at $\sim 1000\,{\rm cm}^{-1}$ is
consistent with a predicted feature at $4\,\Delta_0$ arising from
inelastic scattering from spin fluctuations, and the smooth onset of
this peak is also a natural consequence of a gapless state.

The fact that this same model provides a fit to the microwave
penetration depth and a qualitative fit to the real part of the
microwave conductivity provides further support for the specific
picture of a $d_{x^2-y^2}$ gap with an underlying spin-fluctuation dynamic
interaction.

\acknowledgments

PJH is grateful
for enlightening discussions with D. B. Tanner and W. O. Putikka.
This work was supported by the University of Tennessee and the
Division of Materials Sciences, U.S. Department of Energy, under
Contract No.~DE--AC05--84OR21400 with Lockheed Martin Energy Systems (SMQ),
and by NSF under grant DMR92--25027 (DJS).  The numerical calculations
reported in this paper were performed at the San Diego Supercomputer
Center.

\appendix
\section*{Approximate evaluation of frequency-dependent conductivity}

Here we give an approximate form of the conductivity in the current
model for a $d_{x^2-y^2}$ state suitable for ready comparison with
experimental data, requiring only a single numerical quadrature. The
exact form of the conductivity for a flat band in the absence of
vertex corrections is
\begin{eqnarray}
\sigma_{\perp} (q=0,\omega;T)=-{ne^2\over m} \int d\omega^\prime
W(\omega^\prime)S_\perp (\omega^\prime),
\label{sigma_perp}
\end{eqnarray}
where $W(\omega^\prime;\omega)\equiv [f(\omega^\prime-\omega)-
f(\omega^\prime)]/\omega$ and
\begin{eqnarray}
S_{\perp}(\omega^\prime;\omega)= \mathop{\rm Im}
\int {{d\phi}\over{2\pi}} \cos^2\phi
\Biggl[{{\tilde \omega_+}^\prime\over{{\tilde \omega_+}-
{\tilde \omega_+}^\prime}}\Bigl({1\over{\xi_{0+}}^\prime}-
{1\over{\xi_{0+}}}\Bigr)+{{\tilde \omega_-}^\prime\over{{\tilde
\omega_+}-
{\tilde \omega_-}^\prime}}\Bigl({1\over{\xi_{0+}}}+
{1\over{\xi_{0-}}^\prime}\Bigr) \Biggr].
\label{s_perp}
\end{eqnarray}
Here we have defined
$\tilde\omega_{\pm}=\tilde\omega_{\pm}(\omega^\prime) =
\omega^\prime-\Sigma_0(\omega^\prime\pm i0^+)$,
$\xi_{0\pm}=\xi_\pm(\omega^\prime) =
\pm\mathop{\rm sgn}\omega^\prime\sqrt{{\tilde\omega}_{\pm}^2-\Delta_k^2}$,
as well as analogous primed quantities
$\tilde\omega_{\pm}^\prime=\tilde\omega_{\pm}(\omega^\prime-\omega)$
and $\xi_{0\pm}^\prime=\xi_{0\pm}(\omega^\prime-\omega)$.
In general, Eq.~(\ref{s_perp}) is a complicated expression to evaluate
due to the dependence of $\Sigma_0$ on ${\tilde\omega}$ and the
consequent necessity of evaluating the renormalized frequencies
self-consistently ``in parallel'' with the integral. However in a
sufficiently clean system, self-consistency corrections are important
only over an energy range of negligible interest for the infrared
conductivity.  We therefore make the replacement
${\tilde\omega}_\pm\rightarrow \omega^\prime\pm i/2\tau(\omega^\prime)$,
where $1/\tau(\omega)$ takes the approximate clean limit forms
\begin{eqnarray}
1/\tau(\omega)\simeq \left\{
  \begin{array}{l@{\qquad}l}
    \min\left(0.6\sqrt{\Gamma\Delta_0},\Gamma\Delta_0/|\omega|\right)
                              & |\omega|<\omega^* \\
    0.08\,\Delta_0\,\left(|\omega|/\Delta_0\right)^3
                              & \omega^*<|\omega|<3\,\Delta_0 \\
    0.7|\omega|               & |\omega|>3\,\Delta_0
  \end{array}
\right.
\end{eqnarray}
in the unitary limit and
\begin{eqnarray}
1/\tau(\omega)\simeq  \left\{
  \begin{array}{l@{\qquad}l}
    \Gamma_N|\omega|/\Delta_0 & |\omega|<\omega^* \\
    0.08\,\Delta_0\,\left(|\omega|/\Delta_0\right)^3
                              & \omega^*<|\omega|<3\,\Delta_0 \\
    0.7|\omega|               & |\omega|>3\,\Delta_0
  \end{array}
\right.
\end{eqnarray}
in the Born limit. The crossover frequency $\omega^*$ is
determined by requiring $1/\tau$ to be continuous.

The kernel $S_\perp$ may then be simplified by retaining $1/\tau$
only in the denominators
of Eq.~(\ref{s_perp}), leaving
{\small
\begin{eqnarray}
S_\perp(\omega^\prime;\omega)\simeq\
{
{\omega(M^\prime-M)-{1\over 2}(1/\tau-1/\tau^\prime)(N^\prime-N)}\over
{2\,\omega^2+{1\over 2}(1/\tau-1/\tau^\prime)^2}
}+{
{\omega(M-M^\prime)-{1\over 2}(1/\tau+1/\tau^\prime)(N^\prime+N)}\over
{2\,\omega^2+{1\over 2}(1/\tau+1/\tau^\prime)^2}
}.
\label{s_perp2}
\end{eqnarray}}
Here $N(\omega)$ is the normalized density of states.  In
order to compare with the full numerical evaluation on a square
lattice, we multiply the exact flat band result for a $d_{x^2-y^2}$
state over a cylindrical Fermi surface with the normalized
tight-binding density of states,
\begin{eqnarray}
N(\omega)\simeq \left\{
  \begin{array}{l@{\qquad}l}
    {2|\omega|\over\pi\Delta_0}
    K\left({|\omega|\over\Delta_0}\right)
    {N_t(\omega)\over N_t(0)}
                       & |\omega|<\Delta_0 \\
    {2\over\pi}
    K\left({\Delta_0\over|\omega|}\right)
    {N_t(\omega)\over N_t(0)}
                       & |\omega|>\Delta_0
  \end{array}
\right.,
\end{eqnarray}
where $K$ is the complete elliptic integral of the first kind.
Similarly, $M(\omega)$ is approximated by
\begin{eqnarray}
M(\omega)\simeq \left\{
  \begin{array}{l@{\qquad}l}
    {2\omega \over \pi\Delta_0}
    K\left({\sqrt{\Delta_0^2-\omega^2}\over\Delta_0}\right)
    {N_t(\omega)\over N_t(0)}
                       & |\omega|<\Delta_0 \\
    0                  & |\omega|>\Delta_0
  \end{array}
\right..
\end{eqnarray}
The tight-binding density of states $N_t(\omega)$ is given by
\begin{eqnarray}
N_t(\omega)={1\over2\pi^2t}
            K\left(\sqrt{1-\left({\omega+\mu\over4t}\right)^2}\right).
\end{eqnarray}

\begin{figure}
\epsfysize=6.5in
\epsfbox{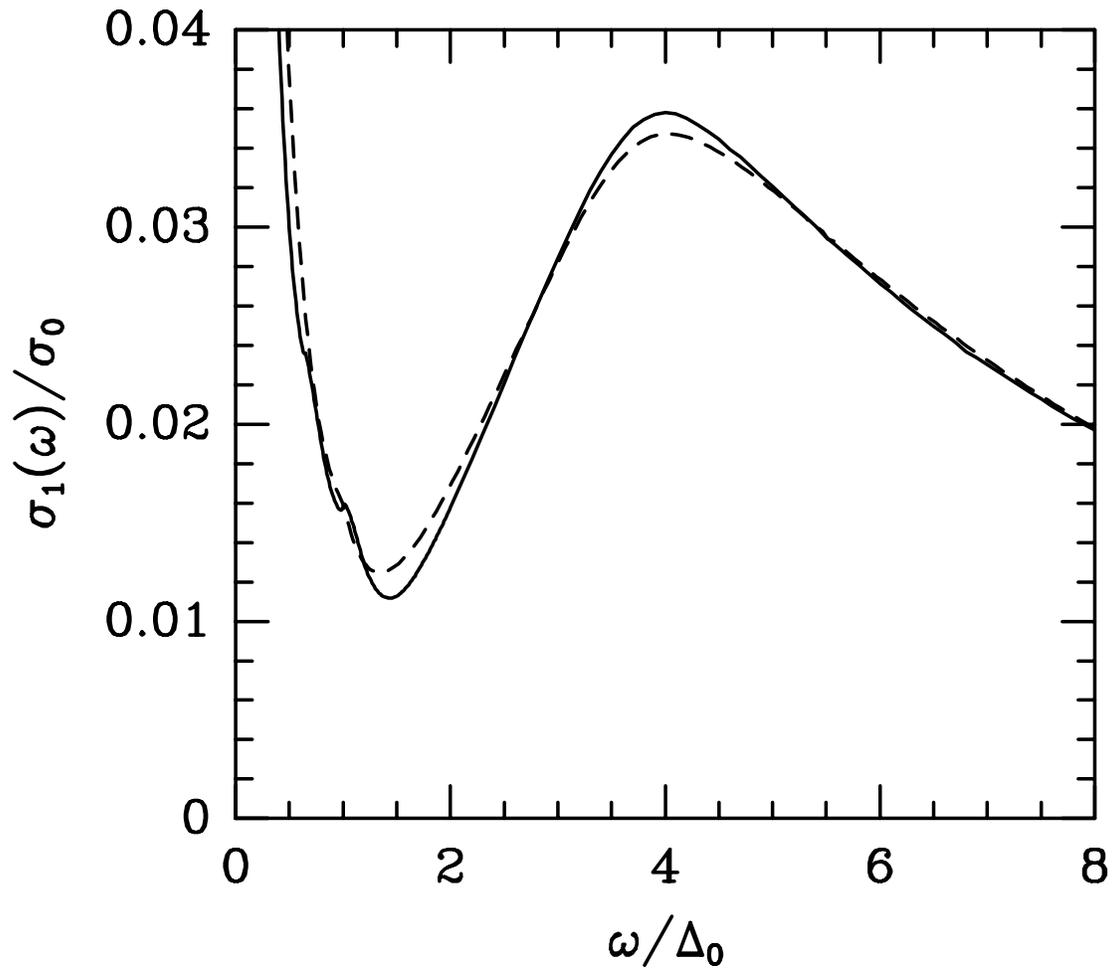}
\caption{A comparison of the results for the real part of the in-plane
conductivity in the superconducting state (for $T=0$ and
$\Gamma=0.018T_c$) obtained from evaluation of Eq.~(\protect\ref{sigma_xx})
(dashed line) and Eq.~(\protect\ref{sigma_perp}) (solid line).}
\label{sig1comp}
\end{figure}

In Eq.~(\ref{s_perp2}), unprimed quantities $M$, $N$, and $1/\tau$ are
evaluated at $\omega^\prime$, whereas primed quantities are evaluated
at $\omega^\prime-\omega$. Substitution into Eq.~(\ref{sigma_perp})
yields an approximation which compares well to the full numerical
evaluation, with an error of only a few percent, as shown in
Fig.~\ref{sig1comp}.

Note that the correct high-frequency limit of the conductivity
in the case of inelastic scattering is {\it not} given by
the usual Drude expression with $1/\tau=1/\tau(\omega)$.  For
example, in the case of a flat band and $1/\tau(\omega)=b\omega$,
the correct high-frequency limit obtained from Eqs.~(\ref{sigma_perp})
and (\ref{s_perp}) is
\begin{eqnarray}
\sigma_1\rightarrow\left({ne^2\over m}\right){b/(2\omega)\over 1+b^2/4},
\end{eqnarray}
not $\sigma_1\rightarrow(ne^2/m)[(b/\omega)/(1+b^2)]$.

\end{document}